\documentclass[10pt]{iopart}

\usepackage{graphicx}

\begin{document}

\title[Validation of entanglement purification by continuous 
variable polarization]
{Validation of entanglement purification by continuous variable
polarization}

\author{Francesca Alessandrini,
Stefano Mancini
\footnote[3]{To
whom correspondence should be addressed (stefano.mancini@unicam.it)},
and Paolo Tombesi 
}

\address{INFM, Dipartimento di Fisica,
Universit\`a di Camerino,
I-62032 Camerino, Italy
}

\begin{abstract}
We investigate the possibility of characterizing 
two-party entanglement
by measuring correlations of Stokes operators in polarized 
bright light beams. We adapt a general separability criterion
to such operators. We then show that entanglement
purification can only be singled out for a
particular protocol. 
\end{abstract}


\submitto{\JOB}

\maketitle

\section{Introduction}

Two-party entanglement of pure states is well characterized by
the von Neumann entropy of one of the two parties \cite{Ben96}.
However, this implies the complete knowledge of the 
system state, i.e. measuring a large set of observables \cite{Tom99}. 
For practical purposes it is preferable
to use quantum correlations of few observables.
Criteria for the inseparability of continuous variable systems
usually refer to two quadratures measurements \cite{Rei88}.
Nevertheless, polarization correlations might be used 
as well to single out entanglement.

The polarization state of light has been extensively studied in the
quantum mechanical regime of single photons.  The
demonstration of entangled polarization states for pairs of
photons has been of particular interest.  This entanglement has
facilitated the study of many interesting quantum phenomena such as
Bell's inequality \cite{Asp82}.  Over the last decade, research on
quantum polarization properties of intense light fields was also 
developed \cite{Kor96}. 
More recently, this topic has attracted attention due 
to the possibility of
transferring continuous variable quantum information
from optical polarization states to the spin state of atomic ensembles
\cite{Hal99}, and to the possibility of local oscillator-free
continuous variable quantum communication networks \cite{Kor02}.  
Some papers have now been published 
which discuss the concept of continuous
variable polarization entanglement, propose methods for its
generation, and provide its experimental evidence
\cite{Kor02, Bow02}.  

The aim of this work is to characterize two-party
entanglement through polarization (Stokes) operators correlations 
and to use them to validate entanglement manipulation (purification).
We begin by discussing, in Section 2, 
continuous variable polarization entanglement
by using a simple separability criterion.
We then analyze the performance of two entanglement 
purification protocols in Section 3 and 
final remarks are outlined in Section 4.

\section{Polarization entanglement}

The polarization state of a light beam can be described as a Stokes
vector on a Poincar\'{e} sphere and is determined by the four Stokes
operators \cite{Jau76}: $\hat S_{0}$ represents the beam
intensity whereas $\hat S_{1}$, $\hat S_{2}$, and $\hat S_{3}$
characterize its polarization and form a cartesian axis system.  
Quasi-monochromatic laser light is almost completely
polarized, and all Stokes operators
can be measured with simple experiments \cite{Kor02}.  Following
\cite{Jau76} we expand the Stokes operators in terms of the
annihilation $\hat a$ and creation $\hat a^{\dagger}$ operators of the
horizontally ($H$) and vertically ($V$) polarized modes
\begin{eqnarray}
\hat S_{0}&=& 
\hat a_{H}^{\dagger} \hat a^{ }_{H} + \hat a_{V}^{\dagger}\hat a_{V}\,,
\qquad\quad
\hat S_{1}= 
\hat a_{H}^{\dagger} \hat a_{H}^{ } - \hat a_{V}^{\dagger}\hat a_{V}^{ }\,,
\label{S0S1}\\
\hat S_{2}&=&
\hat a_{H}^{\dagger} \hat a_{V}^{ } e^{i\theta} \!+ \hat a_{V}^{\dagger}
\hat a_{H}^{ } e^{-i\theta}\,,
\quad
\hat S_{3}=
i\hat a_{V}^{\dagger} \hat a_{H}^{ } e^{-i\theta} \!-i\hat a_{H}^{\dagger} 
\hat a_{V}^{ } e^{i\theta}\,,
\label{S2S3}
\end{eqnarray}
where $\theta$ is the phase difference between the $H$-, $V$-polarization
modes. The commutation relations of the
annihilation and creation operators $[ \hat a_{j} \!  , \!  \hat
a_{k}^{\dagger}] \!  = \!  \delta_{jk}$ with $j,k \!  \in \!\{H,V\}$
directly result in Stokes operator commutation relations,
\begin{equation}
[\hat S_{1}, \hat S_{2}] =  2 i \hat S_{3}\,,\quad
[\hat S_{2}, \hat S_{3}] =  2 i \hat S_{1}\,,\quad
[\hat S_{3}, \hat S_{1}] =  2 i \hat S_{2}\,.
\label{Scr}
\end{equation}
These commutation relations dictate uncertainty relations 
which indicate that entanglement is possible between
the Stokes operators of two beams
(namely, it comes out from their correlations),
and this is termed continuous variables
polarization entanglement.  Three observables are involved, compared
to two for quadrature entanglement, and the entanglement between two
of them relies on the mean value of the third.  

\begin{figure}
\begin{center}
\includegraphics[width=0.6\textwidth]{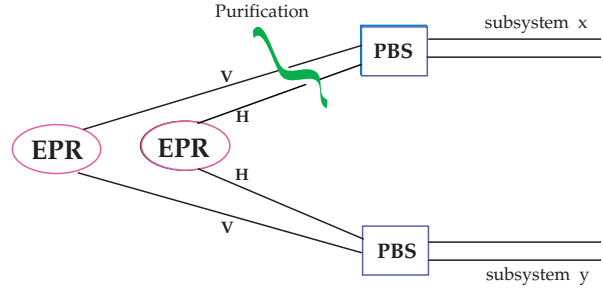}
\end{center}
\vspace{-0.5cm} \caption{\label{fig1} 
Scheme for transforming quadrature entanglement into continuous 
variable polarization entanglement. 
Two quadrature entangled pairs, with different polarizations $H$ and $V$, 
are sent to polarizing beam splitters (PBS).
The emerging beams are then used to measure 
Stokes operators in the two subsystems $x$ and $y$.
Local operations on the subsystem $x$, 
prior mode mixing at PBS, 
allow for entanglement purification. 
}
\end{figure}

The relation between quadrature entanglement and 
polarization entanglement can be understood with the aid of 
fig.~\ref{fig1}.
Two quadrature entangled pairs, with different polarizations $H$ and $V$, 
are sent to two polarizing beam splitters (PBS).
The emerging beams are then used to measure the
Stokes operators (\ref{S0S1}), (\ref{S2S3}) 
in the two subsystems $x$ and $y$.
Local operations on the subsystem $x$, 
before the mode mixing at PBS, 
allow for entanglement purification. 

We assume that the two horizontally and the two vertically
polarized inputs are
intense entangled pairs with fluctuations described by
two-mode squeezed vacuum states \cite{Wal95}
\begin{equation}
|\psi\rangle_{H_{x}H_{y}V_{x}V_{y}}=
\sum_{n=0}^{\infty}
c_{n}|n,n\rangle_{H_{x}H_{y}}
\sum_{m=0}^{\infty}c_{m}|m,m\rangle_{V_{x}V_{y}}\,,
\label{psiHV}
\end{equation}
where $c_{n}=\lambda^n\sqrt{1-\lambda^{2}}$
with $\lambda$ the two-mode squeezing parameter \cite{Wal95}
(for the sake of simplicity we assume $\lambda$ equal for both pairs).
If we introduce the amplitude and phase 
quadrature operators 
\begin{equation}
{\hat X}^{+}_{j}={\hat a}_{j}+{\hat a}^{\dag}_{j}\,,
\qquad
{\hat X}^{-}_{j}=-i\left({\hat a}_{j}-{\hat a}^{\dag}_{j}\right)\,,
\qquad
(j=H\,,\,V)\,,
\label{quad}
\end{equation}
it is easy to see \cite{Wal95}
that for $\lambda\to 1$ each pair of eq.(\ref{psiHV}) tends to be a 
maximally entangled state, 
like the EPR state \cite{Ein35},
for it results simultaneous eigenstate of the difference of  
amplitude quadrature fluctuations 
$\delta{\hat X}^{+}_{j_{x}}-\delta{\hat X}^{+}_{j_{y}}$ ($j=H,V$)
and of the sum of phase quadrature fluctuations
$\delta{\hat X}^{-}_{j_{x}}+\delta{\hat X}^{-}_{j_{y}}$
(we have used the notation $\hat O = \langle \hat O
\rangle + \delta \hat O$). 
For this reason such entanglement is usually refered 
as quadrature entanglement.

Since we have assumed bright input beams, the Stokes operators 
(\ref{S0S1}) and (\ref{S2S3}) 
can be rewritten as
\begin{eqnarray}
{\hat S}_{0}&=&\alpha_{H}^{2}+\alpha_{V}^{2}+\alpha_{H}\,
\delta {\hat X}^{+}_{H}\,,
\label{S0}\\
{\hat S}_{1}&=&\alpha_{H}^{2}-\alpha_{V}^{2}+\alpha_{H}\,
\delta {\hat X}^{+}_{H}-\alpha_{V}\,\delta {\hat X}^{+}_{V}\,,
\label{S1}\\
{\hat S}_{2}&=&2\alpha_{H}\alpha_{V}\cos\theta
+\alpha_{H}\cos\theta\,\delta{\hat X}^{+}_{V}
-\alpha_{H}\sin\theta\,\delta{\hat X}^{-}_{V}
\nonumber\\
&&+\alpha_{V}\cos\theta\,\delta{\hat X}^{+}_{H}
-\alpha_{V}\sin\theta\,\delta{\hat X}^{-}_{H}\,,
\label{S2}\\
{\hat S}_{3}&=&-2i\alpha_{H}\alpha_{V}\sin\theta
-i\alpha_{H}\sin\theta\,\delta{\hat X}^{+}_{V}
-i\alpha_{H}\cos\theta\,\delta{\hat X}^{-}_{V}
\nonumber\\
&&-i\alpha_{V}\sin\theta\,\delta{\hat X}^{+}_{H}
+i\alpha_{V}\cos\theta\,\delta{\hat X}^{-}_{H}\,,
\label{S3}
\end{eqnarray}
where $\alpha_{j}=\langle a_{j} \rangle$ ($j=H,V$).

To provide a proper
definition of the entanglement in terms of the operators
(\ref{S0})-(\ref{S3}), 
we use the general inseparability criterion proposed in 
\cite{Gio03a}. Namely, starting from a generic couple of 
observables $\hat A$ and $\hat B$ for each subsystem
with ${\hat{\cal C}}=i\left[{\hat A},{\hat B}\right]$, 
we construct the following observables on the total system
\begin{eqnarray}
{\hat{\cal U}}&=&r_{x}{\hat A}_{x}+r_{y}{\hat A}_{y}\,,
\qquad r_{x},r_{y}\in{\bf R}\,,
\label{Ucal}
\\
{\hat{\cal V}}&=&s_{x}{\hat B}_{x}+s_{y}{\hat B}_{y}\,,
\qquad s_{x},s_{y}\in{\bf R}\,.
\label{Vcal}
\end{eqnarray}
Then, a sufficient condition for inseparability reads
\begin{equation}
\Delta^{2}{\hat{\cal U}}
+\Delta^{2}{\hat{\cal V}}
<\left(\left|r_{x}s_{x}\right|\,
\left|\langle{\hat{\cal C}}_{x}\rangle\right|
+\left|r_{y}s_{y}\right|\,
\left|\langle{\hat{\cal C}}_{y}\rangle\right|
\right)\,,
\label{ineqUV}
\end{equation}
where $\Delta^{2} \hat O  = \langle
\delta \hat O ^{2} \rangle$.
For the sake of simplicity we choose $r=\pm 1$ and $s=\pm 1$,
then eq.(\ref{ineqUV}) becomes
\begin{equation}
\Delta_{x \pm y}^{2}\hat A+\Delta_{x \pm y}^{2}\hat B 
< 2 \left|\langle\left[\delta \hat A, \delta \hat B\right]\rangle\right|\,.
\label{ineqAB}
\end{equation}
Here $\Delta \!  _{x \pm y}^{2} \!  \hat
O$ is the smaller of the sum and difference variances of the operator
$\hat O$ between beams $x$ and $y$, i.e. $\Delta \!  _{x \pm y}^{2} \! 
\hat O \!  = \!  \min{\langle ( \delta \hat O_{x} \!  \pm \!  \delta
\hat O_{y})^{2} \rangle}$.  
To allow direct analysis of our results, we define the
degree of inseparability $I_{{\hat A},{\hat B}}$, normalized such that
$I_{{\hat A},{\hat B}} \!  < \!  1$ guarantees the state is inseparable
\begin{equation}
I_{{\hat A},{\hat B}}=\frac{\Delta\!  _{x \pm y}^{2}\!  \hat A+\Delta\! 
_{x \pm y}^{2}\!  \hat B} {2 |\langle[\delta \hat A, 
\delta \hat B]\rangle |}
\label{insep}
\end{equation}

Following Ref.~\cite{Bow02}, we arrange the entanglement such that
the mean value of the
three Stokes operators are the same ($| \!  \langle \hat S_{i}\rangle
\!  | \!  = \!  \alpha^{2}$).
This leads to $\alpha_{V}^{2} \!  = \!  \frac{\sqrt{3} -
1}{2}\alpha^{2}$, $\alpha_{H}^{2} \!  = \!  \frac{\sqrt{3} +
1}{2}\alpha^{2}$, $\theta_{x} \!  = \!  \pi/4 \!  + \!  n_{x}\pi/2$,
and $\theta_{y} \!  = \!  \pi/4 \!  + \!  n_{y}\pi/2$ where $n_{x}$
and $n_{y}$ are integers.  
By virtue of eq.(\ref{psiHV}), the two horizontally
polarized inputs, and the two vertically polarized inputs, are
quadrature entangled with the same degree of correlation such that
$\Delta \!  _{x \pm y}^{2} \!  \hat X \!  _{H}^{\pm} \!  = \!  \Delta
\!  _{x \pm y}^{2} \!  \hat X \!  _{V}^{\pm} \!  = \!\Delta \!  _{x
\pm y}^{2} \!  \hat X$.  In this configuration, from
eqs.~(\ref{S0})-(\ref{S3}) and (\ref{psiHV}) one also has 
$|\langle[\delta\hat S_{i}\delta\hat S_{j}]\rangle|=2\alpha^{2}$, 
for all $i\neq j$.  To simultaneously minimize all three degrees of Stokes
operator inseparability ($I_{{\hat S_{i}}, {\hat S_{j}}}$) it is necessary
that $\theta_{x} \!  = \!  -\theta_{y} \!  + \!  n\pi$.  After making
this assumption we find that $\Delta \!  _{x \pm y}^{2} \!  \hat S_{i}
\!  = \!  \sqrt{3}\alpha^{2}\Delta \!  _{x \pm y}^{2} \!  \hat X$ for
all $i$.  Hence, in this situation $I_{{\hat S_{i}}, {\hat S_{j}}}$ are all
identical, and for any pair of Stokes operators the entanglement is the same,
that is  
\begin{equation}
I_{{\hat S}_{i},{\hat S}_{j}}(\lambda)=\sqrt{3}\left(
\frac{1-\lambda}{1+\lambda}\right)\,,
\qquad \forall i\ne j\,.
\label{ISS}
\end{equation}
In principle it is possible to have all
the three Stokes operators perfectly entangled.  In other word,
ideally the measurement of any Stokes operator of one of the beams,
could allow the exact prediction of a different Stokes operator of the
other beam (see fig.~\ref{fig1}).  

\begin{figure}
\begin{center}
\includegraphics[width=0.5\textwidth]{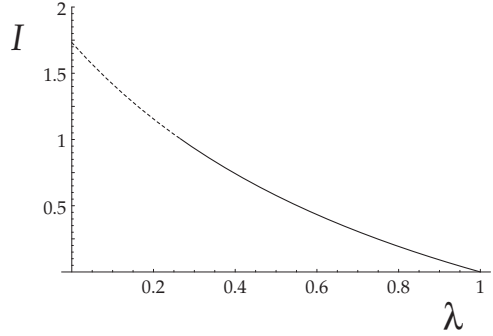}
\end{center}
\vspace{-0.5cm} \caption{\label{fig2} 
Degree of inseparability $I_{{\hat S}_{i},{\hat S}_{j}}(\lambda)$ 
vs $\lambda$.
The dashed part of the line represents values above $1$.
}
\end{figure}

In fig.~\ref{fig2} it is shown the degree of inseparability
(\ref{ISS}) versus the two-mode squeezing parameter $\lambda$.
It is worth noting that entanglement can only be recognized for 
approximately $\lambda > 0.27$ while the state is entangled 
for any values $\lambda\ne 0$. That is, the Stokes operators are not 
optimal entangled witnesses, as they are not tangent to the 
set of separable states \cite{Eck03}. 
Let us now consider the possibility to render the entanglement 
more ``visible" (through polarization correlations) 
for low values of $\lambda$.

\section{Entanglemet Purification}

One important concept in quantum information theory 
is the entanglement purification (distillation) 
which allows the two parties to extract a small number of 
highly entangled, almost pure, states from a large number of weakly 
entangled mixed states \cite{Ben96}. These protocols involve only local 
operations and classical communication (LOCC) between the two parties; 
therefore they can be performed after the distribution of the 
entangled states (see fig.~\ref{fig1}).

Since we are dealing with continuous variable, hereafter we shall consider 
the two relevant protocols developed till now \cite{Dua00,Fiu03}
which we label A and B respectively.
They both involve nonlinear processes; as matter of fact
it was recently proved the impossibility to purify Gaussian 
entangled states by means of Gaussian operations \cite{Eis02},

\subsection{Scheme A}

The scheme proposed by Duan {\it et. al.} \cite{Dua00} relies on 
nondemolition measurement of the total photon number in 
one of the two parties and represents a direct extension of the 
Schmidt projection method to infinite-dimensional Hilbert space.
In this case, the nonlinearity required to implement a non-Gaussian 
transformation is induced by a measurement that should resolve the 
number of photons in one subsystem ($x$ in fig.~\ref{fig1}). 

Suppose the result of the measurement is $J$, then the state after 
measurement will be 
\begin{equation}
|J\rangle_{H_{x}H_{y}V_{x}V_{y}}=
\frac{1}{\sqrt{J+1}}\sum_{n=0}^{J}
|n,n\rangle_{H_{x}H_{y}}|J-n,J-n\rangle_{V_{x}V_{y}}\,,
\label{JHV}
\end{equation}
and the probability for the random outcome $J$
can be calculated from eqs.(\ref{psiHV}) and (\ref{JHV}) as
\begin{equation}
\left|\langle\psi|J\rangle\right|^{2}=
(J+1)\lambda^{2J}(1-\lambda^{2})^{2}\,.
\label{psiJ}
\end{equation}
Then, the degree of inseparability (\ref{insep}) calculated on the 
state (\ref{JHV}) gives
\begin{equation}
I_{{\hat S}_{i},{\hat S}_{j}}(J)
=\frac{\sqrt{3}}{J+1}\left[
\sum_{m=0}^{J}\left(2m+1\right)
-\sum_{m=0}^{J-1}\sqrt{\left(J-1\right)\left(m+1\right)}\right]\,,
\label{IJ}
\end{equation}
which results independent of $\lambda$.
\begin{figure}
\begin{center}
\includegraphics[width=0.45\textwidth]{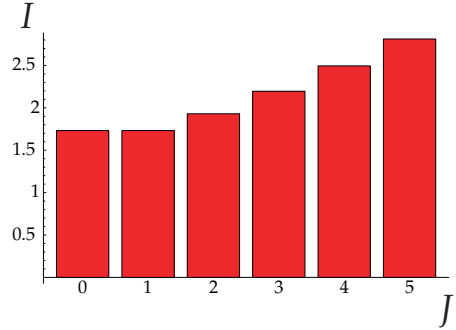}
\end{center}
\vspace{-0.5cm} \caption{\label{fig3} 
Degree of inseparability $I_{{\hat S}_{i},{\hat S}_{j}}(J)$ vs $J$.
}
\end{figure}
Fig.~\ref{fig3} shows that $I_{{\hat S}_{i},{\hat S}_{j}}(J)$
never goes below $1$ (whatever is the value of $\lambda$).
Hence the purification protocol does not
succeed in improving the visibility of entanglement on 
polarization correlations,
although the entanglement has been effectively enhanced \cite{Dua00}.

\subsection{Scheme B}

The scheme proposed by Fiurasek \cite{Fiu03} provides
entanglement purification for any single copy of a two-mode
squeezed vacuum state $\sum_{n}c_{n}|n,n\rangle$.
The procedure in this case preserves the structure of a two-mode 
squeezed vacuum state, 
while the Schmidt coefficients $c_{n}$ are 
transformed to different ones, $c_{n}\to d_{n}$.
The nonlinearity is provided by a cross Kerr interaction 
with an auxiliary mode prepared in a coherent state $|\alpha\rangle$
and undergoing a phase shift $\phi$. 
A subsequent eight-port homodyne detection of the auxiliary mode
provides a random outcome $\beta\in{\bf C}$ 
(projection of the auxiliary mode onto a coherent state $|\beta\rangle$).

In this case the entanglement purification is described through the
replacement 
\begin{equation}
c_{n}\to d_{n}=\frac{\langle\beta|\alpha e^{in\phi}\rangle}
{\sqrt{\pi Q(\beta)}}c_{n}\,,
\label{cndn}
\end{equation}
where $Q(\beta)$ represents the probability density for the outcome
$\beta$, that is
\begin{equation}
Q(\beta)=\frac{1}{\pi}
\sum_{n=0}^{\infty} |\langle\beta|\alpha e^{in\phi}\rangle|^{2}
\, |c_{n}|^{2}\,.
\label{Qbeta}
\end{equation}
The degree of inseparability (\ref{insep})
prior the purification is given by eq.(\ref{ISS}) and can be rewritten as
\begin{equation}
I\left(\lambda\right)
=\sqrt{3}
\sum_{n,m=0}^{\infty}\left[\left(2n+1\right)|c_{n}|^{2}|c_{m}|^{2}
-2\left(n+1\right)|c_{m}|^{2}c_{n+1}c_{n}\right]\,,
\label{Ilam}
\end{equation}
so that after the purification it simply becomes
\begin{equation}
I\left(\beta\right)
=\sqrt{3}
\sum_{n,m=0}^{\infty}\left[\left(2n+1\right)|d_{n}|^{2}|d_{m}|^{2}
-2\left(n+1\right)|d_{m}|^{2}d_{n+1}d_{n}\right]\,.
\label{Ibeta}
\end{equation}
Then, we can introduce the entanglement increment by
\begin{equation}
\Gamma(\beta,\lambda)=
\frac{E'(\beta)}
{E(\lambda)}\,,
\label{Gam}
\end{equation}
where we have set
\begin{equation}
E'(\beta)=
\left\{
\begin{array}{lll}
1-I(\beta)\,,& & {\rm if}\; I(\beta)<1
\\
0\,,& & {\rm otherwise}
\end{array}\right.\,,
\label{Ebeta}
\end{equation}
as the degree of entanglement after the purification and
\begin{equation}
E(\lambda)=
\left\{
\begin{array}{lll}
1-I(\lambda)\,,& & {\rm if}\; I(\lambda)<1
\\
0\,,& & {\rm otherwise}
\end{array}\right.\,,
\label{Elam}
\end{equation}
as the degree of entanglement prior the purification.
Here, $\Gamma >1$ indicates the ability to 
recognize the entanglement improvement through the 
Stokes operators.
The efficiency of the protocol can be defined as \cite{Man01}
\begin{equation}
\Xi(\lambda)=1-\frac{1}{\Upsilon(\lambda)}\,,
\label{Xi}
\end{equation}
where
\begin{equation}
\Upsilon(\lambda)=\frac{\int_{\Omega}
d^{2}\beta\, Q^{2}(\beta)\Gamma(\beta,\lambda)}
{\int_{\Omega}d^{2}\beta\,Q^{2}(\beta)}\,,
\label{Up}
\end{equation}
with $\Omega\equiv\{\beta\in{\bf C}|\Gamma(\beta)>1\}$.
In eq.(\ref{Up}) we have used $Q^{2}$ as probability density because of 
two independent purification processes occur, one for each pair.
\begin{figure}
\begin{center}
\includegraphics[width=0.5\textwidth]{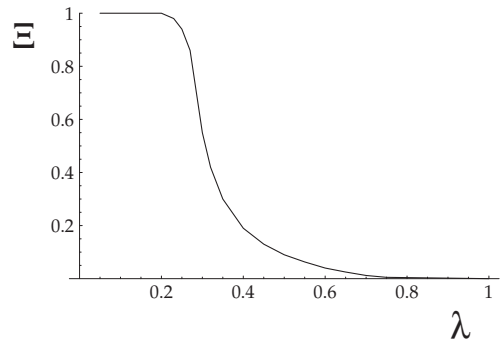}
\end{center}
\vspace{-0.5cm} \caption{\label{fig4} 
Efficiency of the purification protocol B vs $\lambda$.
}
\end{figure}
In fig.~\ref{fig4} we show the efficiency of the protocol 
as function of the parameter $\lambda$.
It takes its maximum for approximately $\lambda < 0.27$
because, prior the purification, the Stokes` operators
are not good enough operators to witness entanglement in this region
of $\lambda$ and the purification protocol is efficient
(see fig.~\ref{fig2}).
Obviously, the graph has a singular point in $\lambda=0$ where
no entanglement is present at all.
For $\lambda > 0.27$ the protocol`s  
efficiency rapidly decreases to zero because
for these values of $\lambda$
the Stokes operators are good enough
to recognize the entanglement present in the 
state (see fig.~\ref{fig2}).

\section{Conclusions}

In conclusion, we have characterized two-party entanglement 
through polarization (Stokes operators) correlations.
Stokes operators turn out to be useful in describing the
transfer of quantum information from a freely propagating 
optical carrier to a matter system \cite{Hal99}.
Furthermore, the use of continuous variable
polarization entanglement combines the advantages of intense, easy to 
handle, sources of EPR-entangled light and efficient direct detection,
thus opening the way to secure quantum communication with bright light
\cite{Ral00}.
However, we have shown that Stokes operators
are not able to single out any level of entanglement from a 
given state. They well work when the degree of entanglement 
is high enough.
Moreover, in order to make entanglement visible 
on polarization correlations, the
purification should be accomplished with a suitable protocol.

The question whether one could ever recognize full entanglement 
with measurements of Stokes operators correlations could be
addressed by optimizing the used entanglement criterion,
or by exploring more sophisticated version of it \cite{Gio03b}.
This is planned for future work.

\section*{Acknowledgements}
S. M. would like to thank Warwick Bowen and Vittorio Giovannetti
for useful comments.

\end{document}